\documentclass[twocolumn,english,aps,prl,reprint,superscriptaddress]{revtex4-1}
\usepackage[LGR,T1]{fontenc}
\usepackage[latin9]{inputenc}
\usepackage{amsmath}
\usepackage{amssymb}
\usepackage{graphicx}
\usepackage[version=4]{mhchem}
\usepackage{xcolor}
\usepackage{makecell}
\usepackage[none]{hyphenat}
\usepackage[english]{babel}
\usepackage{comment}
\usepackage[colorlinks=true, urlcolor=blue, linkcolor=black, citecolor=black]{hyperref}
\usepackage{soul}

\makeatletter

\def\urlprefix{}
\def\url#1{}

\makeatother

\usepackage{babel}
\begin{document}
\title{Majorana Zero Modes Emulated in a Magnetic Molecule Chain}
\author{Silas Hoffman}
\affiliation{Department of Physics, Center for Molecular Magnetic Quantum Materials
and Quantum Theory Project, University of Florida, Gainesville, Florida
32611, USA}
\author{Jie-Xiang Yu}
\affiliation{Department of Physics, Center for Molecular Magnetic Quantum Materials
and Quantum Theory Project, University of Florida, Gainesville, Florida
32611, USA}
\author{Shuang-Long Liu}
\affiliation{Department of Physics, Center for Molecular Magnetic Quantum Materials
and Quantum Theory Project, University of Florida, Gainesville, Florida
32611, USA}
\author{ChristiAnna Brantley}
\affiliation{Department of Chemistry and Center for Molecular Magnetic Quantum Materials,
University of Florida, Gainesville, Florida 32611, USA}
\author{Gautam D. Stroscio}
\affiliation{Division of Chemistry and Chemical Engineering, Arthur Amos Noyes Laboratory of Chemical Physics, California Institute of Technology, Pasadena, California 91125, USA}
\author{Ryan G. Hadt}
\affiliation{Division of Chemistry and Chemical Engineering, Arthur Amos Noyes Laboratory of Chemical Physics, California Institute of Technology, Pasadena, California 91125, USA}
\author{George Christou}
\affiliation{Department of Chemistry and Center for Molecular Magnetic Quantum Materials,
University of Florida, Gainesville, Florida 32611, USA}
\author{Xiao-Guang Zhang}
\affiliation{Department of Physics, Center for Molecular Magnetic Quantum Materials
and Quantum Theory Project, University of Florida, Gainesville, Florida
32611, USA}
\author{Hai-Ping Cheng}
\thanks{Correspond to: hping@ufl.edu}
\affiliation{Department of Physics, Center for Molecular Magnetic Quantum Materials
and Quantum Theory Project, University of Florida, Gainesville, Florida
32611, USA}\begin{abstract}
We propose molecular magnets as a platform to emulate Majorana zero modes (MZMs). Using a quantum chemistry approach, we identify several candidates and predict a Co trimer to have sufficient properties to host MZMs. Parameters of the quantum spin Hamiltonian describing the three coupled magnetic centers are extracted from ab initio calculations. The low-energy subspace of this material realizes an effective anisotropic spin-1/2 chain. We show the presence of MZMs in this system and find that their response to electronic paramagnetic resonance provides an experimentally realizable signature. 
\end{abstract}

\maketitle

Majorana zero modes (MZMs) are persistent exotic states localized to the ends of one-dimensional topological superconductors, and have garnered a particular interest owing to their non-Abelian exchange statistics \cite{ivanovPRL01} and potential application for quantum computing \cite{nayakRMP08,aguadoRT20}. Theoretical models that realize topological superconductivity by constructing heterostructures of conventional materials, such as semiconducting quantum wires and $s$-wave superconductors \cite{lutchynPRL10,oregPRL10,aliceaPRB10}, have launched a fervent experimental search to find MZMs. While the appeal to realize topological superconductivity using conventional materials is attractive, their complicated synthesis and measurement \cite{mourikSCI12,dengNANOL12,dasNATP12,churchillPRB13,albrechtNAT16} can lead to false positives in the identification of MZMs \cite{liuPRB17,reegPRB18,pradaNRP20,yuNATP21}. 

Rather than realizing MZMs in electronic systems, several proposals have suggested emulation of MZMs \cite{jiangPRL11} by exploiting the mapping  \cite{liebAoP61} between a spin-1/2 Ising chain and a spinless one-dimensional topological superconductor supporting localized MZMs \cite{kitaevPU01}. In contrast to semiconductor-based MZMs, the nonlocal transformation which fermionizes the spins also destroys their topological protection. Nonetheless, recent experimental results suggest that MZMs emulated in photonic systems \cite{xuNATC16} and on quantum computers \cite{stengerPRR21} retain their unconventional exchange statistics. While these experimental achievements have shown an impressive control of quantum systems, perhaps a more natural environment to realize MZMs is directly in a quantum spin chain \cite{tserkovnyakPRA11,pedrocchiPRB12}. Moreover, in quantum spin chains, several exotic properties of electronic MZMs, such perfect Andreev reflection \cite{hoffmanCM18}, fractional Josephson current \cite{shenPRR21}, and non-Abelian braiding statistics \cite{backensPRB17}, have a corresponding spin analogue.

Using spin chains as a platform for MZMs faces difficult challenges. On the one hand, the coupling between moments on the chain must be sufficiently strong and anisotropic to ensure an ordered ground state and a gapped bulk spin excitation. On the other hand, weak coupling and low-dimensionality cause disorder, which is detrimental to both the existence of the MZMs and their effective detection. Consequently, neither a concrete experimental realization nor a theoretically proposed material has been suggested to emulate MZMs using a quantum spin chain. These difficulties can be bypassed using magnetic molecules and exploiting the accrued expertise of their field of study. The power of using molecular magnets is in (1) the vast freedom in tuning magnetic anisotropy and inter-metal ion exchange by local chemical environment and by ligands, (2) monodispersive nature assuring low disorder, and (3) ab initio techniques which can quantitatively predict the expected experimental parameters \cite{RN3005, RN3325, RN3327}

In this letter, we suggest one such viable molecular magnet example: \ce{Co3 (SALPN)2 (O2CCH3)2 . (OCHNH2)2}. Using first-principles density functional theory (DFT)~\cite{Hohenberg1964, Kohn1965}, we build a quantum spin model for this candidate. Projecting onto a low-energy subspace of states, we realize a rich effective Hamiltonian which, for a sufficiently small external magnetic field, supports the emulation of MZMs. By simulating an electron paramagnetic resonance (EPR) spectroscopy measurement, we find that the effective occupancy of the MZMs can be extracted and we predict the expected experimental EPR signatures of this material.

\begin{figure}
\includegraphics[width=\columnwidth]{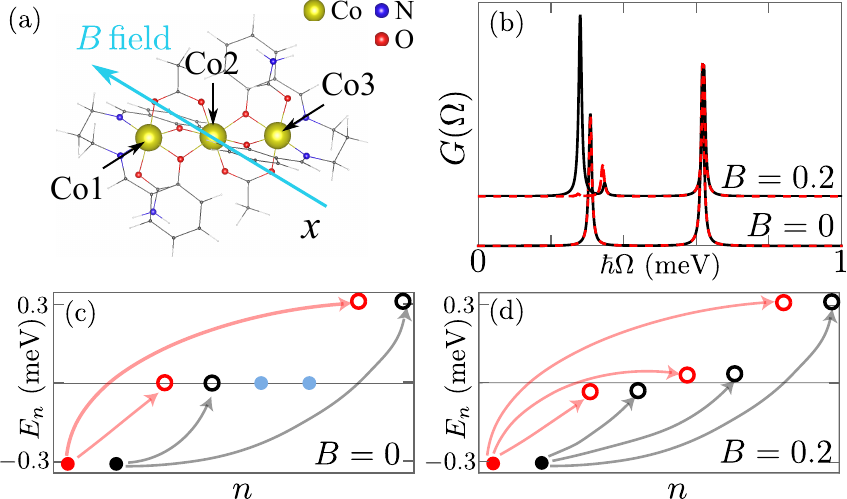}
\caption{\label{fig:co-trimer-struct_eprb} (a) Atomic structure of a single $\textrm{Co}_3$ molecule with $\textrm{OCHNH}_2$ solvent molecules in a dc magnetic field $B$ applied perpendicular to the magnetic easy axes of Co atoms. (b) EPR signal as a function of ac-applied magnetic field frequency $\Omega$. The solid (black) and dashed (red) line are the response of the two ground degenerate states. The spectra are plotted for $B=0$ (c) and $B=0.2$ (d) where ac-magnetic-field-induced transitions between the ground states (filled red circles and black circles) and the accessible excited states (open red and black circles) are indicated by arrows; inaccessible excited states are marked by filled blue circles.}
\end{figure}

\textit{DFT calculations.---} To keep the calculations simple in the search for viable spin chain candidates, we study molecular magnets with three metal centers, which are the smallest molecule chains that can host MZMs. Having studied several candidate materials with sizable anisotropy \cite{SM}, we focus on the Co trimer \ce{Co3 (SALPN)2 (O2CCH3)2 . (OCHNH2)2} which has been experimentally synthesized with a different solvent~\cite{RN3270}. Fig.~\ref{fig:co-trimer-struct_eprb}(a) shows the atomic structure of the Co trimer. The three Co atoms from left to right are labeled as Co1, Co2, and Co3, respectively. Each Co atom has a local spin of $S=3/2$ corresponding to the $+2$ oxidation state. 

This molecule is well-described by a spin Hamiltonian, $\hat H=\hat H_\textrm{ex}+\hat H_a+\hat H_r$, with an isotropic exchange interaction, $\hat H_\textrm{ex}=J\sum^2_{i=1}\hat{\boldsymbol S}_{i}\cdot\hat{\boldsymbol S}_{i+1}$, and an on-site axial and rhombic anisotropy,
\begin{align}
\hat H_a = \sum_{i=1}^3 [ D_i (\hat{\boldsymbol{S}}_{i} \cdot \boldsymbol z_i)^2]\,,
\label{Ha}
\end{align}
and
\begin{align}
\hat H_r = \sum_{i=1}^3  E_i [(\hat{\boldsymbol{S}}_{i} \cdot \boldsymbol x_i)^2 - (\hat{\boldsymbol{S}}_{i} \cdot \boldsymbol y_i)^2]\,,
\end{align}
respectively. $J$ is the exchange coupling constant between two nearest-neighbor Co atoms and $D_i$ and $E_i$ are local axial and rhombic zero field splitting parameters for the $i$th Co atom. The axes of anisotropy for each Co atom are, in general, noncollinear and therefore we express the anisotropy Hamiltonian in the local frame defined by unit vectors $\boldsymbol x_i$, $\boldsymbol y_i$ and $\boldsymbol z_i$. Because of the inversion symmetry of this complex, the exchange coupling is constant and the on-site anistropies of Co1 and Co3 are equal, i.e. $D_1=D_3$ and $E_1=E_3$.

\begin{figure}
\includegraphics[width=0.9\columnwidth]{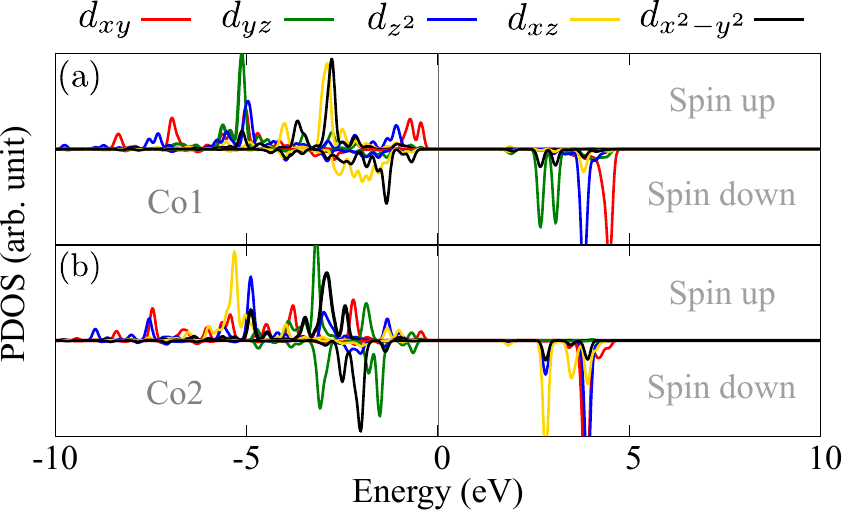}
\caption{\label{fig:co-trimer-pdos}PDOS of the \ce{Co3} molecule with $\textrm{OCHNH}_2$ solvent molecules. The panel (a) is for Co1 at the end of the Co chain and (b) for Co2 in the middle. The Fermi level is set to zero.}
\end{figure}

DFT calculations are carried out using the Vienna Ab initio Simulation Package~\cite{Kresse_1996_CMS, Kresse_1996_PRB} with an energy cutoff of 600 eV for plane waves, the projector augmented wave pseudopotentials~\cite{PAW_1994,PAW_1999} and the Perdew-Burke-Ernzerhof (PBE) exchange-correlation energy functional~\cite{PBE}. The energy and force tolerances are set to $10^{-8}\, \textrm{eV}$ and $0.001\, \textrm{eV/\AA}$ respectively. We fix the magnetic core including Co, O and N atoms to the experimental structure and impose inversion symmetry during structural relaxation. We apply the DFT+$U$ method~\cite{Liechtenstein_1995a} and set the on-site Coulomb and exchange parameters to be $5.0$ and $0.9$ eV respectively for Co $3d$ electrons to account for strong correlation effects. We extract $J$ using the broken-symmetry method~\cite{Ruiz_1999,Ruiz_2003} based on DFT total energies without spin-orbit coupling (SOC). Anisotropy parameters $D_i$ and $E_i$ are extracted by treating spins as classical quantities based on DFT total energies with SOC, which is included self-consistently with constraint on the direction of local spins.~\cite{SLLRN901} The fitted values are $J=-0.092$ meV, $D_1=D_3=-0.668$ meV, $D_2=-0.522$ meV, $E_1=E_3=-0.069$ meV, $E_2=-0.033$ meV. The negative $J$ signifies ferromagnetic coupling between Co1 (or Co3) and Co2. The negative $D_i$ values indicate that each Co atom has a magnetic easy axis. The direction of magnetic easy axis is $(\theta=42.9^\circ, \phi=102.4^\circ)$ for Co1 (or Co3) and $(\theta = 26.5^\circ,  \phi=177.8^\circ)$ for Co2. The difference in the local magnetic anisotropy between Co1 (or Co3) and Co2 originates from the different chemical environment. Co1 (or Co3) is bonded with four oxygen atoms and two nitrogen atoms while Co2 is bonded with six oxygen atoms. As a result, $d$ orbitals are split and occupied differently as shown by the projected density of states (PDOS) in Fig. \ref{fig:co-trimer-pdos}. The ratio $D_i/J$ is 7.3 for Co1 (or Co3) and 5.7 for Co2. 

\textit{Effective Hamiltonian.---} While MZMs are known to be realized in an Ising chain \cite{SM}, the model Hamiltonian used in the DFT calculation is considerably more complicated. Nonetheless, because the exchange interaction and rhombic anisotropy are small compared to the axial anisotropy, we can project onto a subspace of states. Our low energy basis is spanned by the eigenstates of $\hat H_a + \hat H_B$ where the first term is defined by Eq.~(\ref{Ha}) and $\hat H_B=-B\sum_{i=1}^3 \boldsymbol{S}_i\cdot(\boldsymbol z_1\times\boldsymbol z_2)$ describes an applied magnetic field simultaneously perpendicular to the easy axes of the Co atoms. We label the low energy eigenstates of each Co site by $\left|\uparrow\right\rangle$ and $\left|\downarrow\right\rangle$, respectively. Each site effectively furnishes two states, analogous to the spin-1/2 chain, and the tensor product of these states furnish a low energy basis.

Projecting $\hat H+\hat H_B$ onto the low energy states, we find the effective Hamiltonian takes the form,
\begin{align}
\hat H^\textrm{eff}=&-\sum^2_{i=1}\left[J^x\sigma^x_i\sigma^x_{i+1}+J^y\sigma^y_i\sigma^y_{i+1}+J^z\sigma^z_i\sigma^z_{i+1}\right.\nonumber\\
&\left.+\mathcal D (\sigma_i^x\sigma^y_{i+1}-\sigma_i^y\sigma^x_{i+1})\right]-\sum_{i=1}^3 \mu_i \sigma^z_i\,,
\label{XYZ}
\end{align}
where the Pauli matrices act on the low energy basis of the $i$th Co. Eq.~(\ref{XYZ}) is a spin-1/2 Heisenberg XYZ model with a Dzyaloshinskii-Moriya interaction (DMI) and site-dependent magnetic field. The spin-1/2 Hamiltonian inherits the on-site anisotropy of the spin-3/2 model as an anisotropy in the exchange interaction, i.e. $J_x\neq J_y\neq J_z$ in general. The noncollinearity of the axes of anisotropy is expressed as a renormalization of the exchange interaction parameters as well as a Dzyaloshinskii-Moriya interaction. This broken inversion symmetry, although not present in the original spin-3/2 Hamiltonian, is necessary to capture the effect of noncollinear anisotropy which will manifest in the EPR spectra below.

The parameters have an implicit, albeit convoluted, dependence on the magnetic field and we have dropped a magnetic-field-dependent constant energy. When the magnetic field is small compared to the axial anisotropy, $b=B/D\ll 1$, the parameters entering Eq.~(\ref{XYZ}) are polynomial in $b$,
\begin{align}
J^x&=J S^2(1-b^2/4 -S b^4/16)\cos\theta\,,\nonumber\\
J^y&=(J S^4 b^4/16)\cos\theta\,,\,\,\,\,J^z=J S^4 b^4/16\,,\nonumber\\
\mathcal D &= J (b^2 /16 - S^2 b^4 /16)\sin\theta\,,\nonumber\\
\mu_1&=\mu_3= S(b E-b^3[4(1+E)+9J)/32\,,\nonumber\\
\mu_2&= S(b E-b^3[2(1+E)+9J)/16\,,
\label{Heff}
\end{align}
where, for simplicity, we have considered $D_1=D_2=D_3=D$ and $E_1=E_2=E$. $\theta$ is the angle between the easy axes, i.e. $\boldsymbol z_1\cdot\boldsymbol z_2=\cos\theta$. 

\begin{figure}[t!]
\includegraphics[width=\columnwidth]{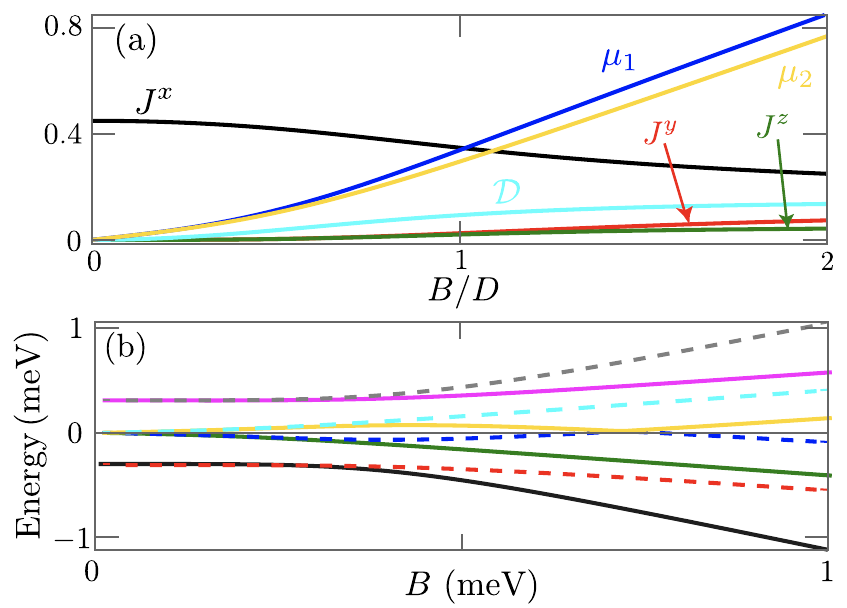}
\caption{(a) Values of effective parameters in Eq.~(\ref{XYZ}), in units of $D_1=D_2=D_3=D$, obtained by projecting the trimer Hamiltonian with $J=0.2D$ and $E_1=E_2=0.1D$ and $\theta=0$. (b) Spectrum of Eq.~(\ref{XYZ}) as a function of applied magnetic field, $B$, approximately along the $(\theta,\phi)\approx(\pi/3,-\pi/10)$, such that it is perpendicular to the axes of anisotropy of the Co's, using the parameters found using DFT.}
\label{params}
\end{figure}

When $B=0$, Eq.~(\ref{XYZ}) is a spin-1/2 Ising model wherein the MZMs are written as $\gamma=\sigma^x_1$ and $\gamma'=\sigma_1^z\sigma_2^z\sigma_3^y$ \cite{SM}. Plotting the parameters in Eq.~(\ref{XYZ}) for homogeneous exchange and anisotropy and similar $J/D$ and $E/D$ ratios to those obtained in the DFT calculation [Fig.~\ref{params}(a)], we find that the Ising exchange largely dominates over the exchange interaction and DMI. Accordingly, we estimate the topological phase transition to be roughly when $\mu_2\simeq J^x$, corresponding to $B\approx D$. That is, for an infinite chain, MZMs remain exponentially localized to the ends of the chain when $B<D$ and vanish when $B>D$. Employing the methodology outlined in Ref.~\cite{alexandrandinataPRB16}, one can find an explicit expression for the MZMs at finite $B$ dressed according to the additional interactions in Eq.~(\ref{XYZ}).

With the effective Hamiltonian prepared, upon allowing for inhomogeneous axial and rhombic anisotropy, we proceed to calculate the spectrum using the values obtained from DFT [Fig.~\ref{params}(b)]. The presence of MZMs is reflected in the double degeneracy of the states. When $B=0$,  the system is simply an Ising chain and the lowest doublet of states correspond to fully spin polarized states, $\left|\uparrow\uparrow\uparrow\right\rangle$ and $\left|\downarrow\downarrow\downarrow\right\rangle$. The four states at zero energy are those states with single spin flip at the end of the chain, $\left|\downarrow\uparrow\uparrow\right\rangle$, $\left|\uparrow\uparrow\downarrow\right\rangle$, $\left|\uparrow\downarrow\downarrow\right\rangle$, and $\left|\downarrow\downarrow\uparrow\right\rangle$. The highest energy correspond to the middle spin being flipped, $\left|\uparrow\downarrow\uparrow\right\rangle$ and $\left|\downarrow\uparrow\downarrow\right\rangle$.
As the magnetic field is turned on, these states mix according to the effective Hamiltonian. The MZMs overlap and begin to hybridize when $B\approx0.3D$, breaking the double degeneracy. This dependence of the spectrum on the magnetic field can be used as a signature that MZMs are present in the trimer.

\textit{Spectral analysis.---} To probe the spectrum, we calculate the response of a ground state to a transverse ac-magnetic field, i.e. simulate EPR. The ac-magnetic field within the spin-3/2 model is $\hat H_\textrm{EPR}(t)=h_\textrm{ac}\cos(\Omega t)\sum_{i=1}^3 \boldsymbol{S}_i\cdot(\boldsymbol z_1\times\boldsymbol z_2)$ which maps to $\hat H_\textrm{EPR}^\textrm{eff}(t)=\cos(\Omega t)\sum_{i=1}^{3} h_{i,\textrm{ac}}^\textrm{eff}\sigma_i^z$ with $h_{i,\textrm{ac}}^\textrm{eff}=SB/4D_i$ in the effective spin-1/2 model. The linear response of the system to the ac field is \cite{berimJETP79,kuboJoPSoJ54}
\begin{equation}
G(\Omega)\sim\int \textrm{d}t\cos(\Omega t)\left\langle \biggl\{ \sum_{i=1}^{N}h_{i,\textrm{ac}}^\textrm{eff}\sigma_{i}^{z},\sum_{j=1}^{N}h_{j,\textrm{ac}}^\textrm{eff}\sigma_{j}^{z}(t)\biggr\} \right\rangle \,,\label{resp}
\end{equation}
where the expectation is with respect to one of the ground states
and $\sigma^{z}_j(t)$ evolves according to Eq.~(\ref{XYZ}). The form of the ac-magnetic field implies that the resultant EPR signal measures transitions between states that differ by one spin flip; this statement is exact in the case when $B=0$ and can be used to guide our intuition for finite magnetic field.

When $B=0$, the EPR signal [Fig.~\ref{fig:co-trimer-struct_eprb}(b)] is identical for the two ground states, i.e. the black solid line and red dashed line overlap. This is a consequence of the double degeneracy of the ground states and excited states [Fig.~\ref{fig:co-trimer-struct_eprb}(c)]. Focusing on the positive spin ground state, indicated by the red dot in Fig.~\ref{fig:co-trimer-struct_eprb}(c), the peak at $\Omega\approx0.3$ corresponds to the transition between  $\left|\uparrow\uparrow\uparrow\right\rangle$ and $(\left|\downarrow\uparrow\uparrow\right\rangle+\left|\uparrow\uparrow\downarrow\right\rangle)/\sqrt{2}$ and the peak at $\Omega\approx0.6$ corresponds to  the transition between $\left|\uparrow\uparrow\uparrow\right\rangle$ and $\left|\uparrow\downarrow\uparrow\right\rangle$. We note that the state $(\left|\downarrow\uparrow\uparrow\right\rangle-\left|\uparrow\uparrow\downarrow\right\rangle)/\sqrt{2}$ is inaccessible because it is inversion asymmetric. Reversing the spin direction, one obtains the transitions between the other ground state, $\left|\downarrow\downarrow\downarrow\right\rangle$, and the analogous excited states, highlighted in black and grey in Fig.~\ref{fig:co-trimer-struct_eprb}(c). 

When $B=0.2$, the spectrum [Fig.~\ref{fig:co-trimer-struct_eprb}(d)] shows a splitting of the four states at zero energy to two pairs of doublets while the highest and lowest energy states remain degenerate. While difficult to discern in the spectrum, the near zero energy are not degenerate, i.e. $E_3\neq E_4$ and $E_5\neq E_6$. Moreover, because the effective inversion symmetry is broken in a finite magnetic field, i.e. $\mathcal D\neq0$, the ac-magnetic field induces transitions between one of the ground states and half of the excited states. As a result, the EPR signal displays five peaks [Fig.~\ref{fig:co-trimer-struct_eprb}(b)]. 

\begin{figure}[t!]
\includegraphics[width=\columnwidth]{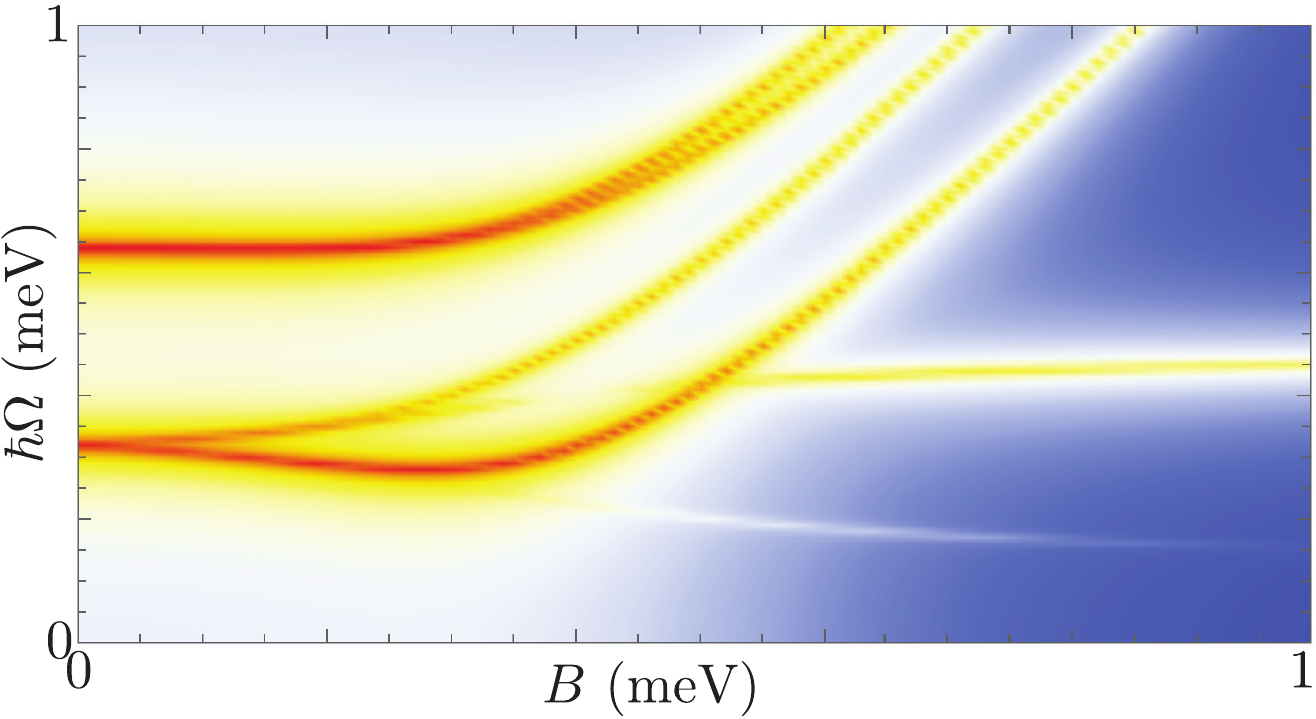}
\caption{EPR spectrum as a function of ac-magnetic field frequency, $\Omega$, and magnitude of dc-applied magnetic field, $B$, approximately along the $(\theta,\phi)\approx(\pi/3,-\pi/10)$, such that it is perpendicular to the axes of anisotropy of the Co's, using the parameters found using DFT. Here, we have plotted the logarithm of the intensity to emphasize the presence of smaller peaks.}
\label{spec2D}
\end{figure}

In Fig.~\ref{spec2D}, we plot the EPR spectrum as a function of $\Omega$ and $B$ by summing the EPR intensities from both of the ground states. Here, we have plotted the logarithm of the intensity in order to highlight the fainter features. Should the molecules be probed in thermal equilibrium at temperatures much smaller than the exchange interaction, we expect to observe a spectral dependence on the magnetic field as seen in Fig.~\ref{spec2D}. 

\textit{Discussion}.--- Despite having a deceptively simple starting point of a spin-$3/2$ Hamiltonian with a homogeneous exchange interaction and on-site anisotropy, the effective spin-$1/2$ Hamiltonian realizes an anisotropic exchange and breaks inversion symmetry. This points to advantages of molecular magnetic materials as candidates for designer effective Hamiltonians. The machinery and methodology in this work can be used as a blueprint to survey molecular magnetic systems to realize a variety of interesting, effective Hamiltonians which support interactions that would be a priori unexpected. For instance, this analysis can be applied to material searches for inhomogenous Ising chains, inhomogenous spin ladders, or Kitaev lattices, all of which support Majorana modes at their ends or edges, respectively. Moreover, upon considering a broader class of materials, we expect such an analysis could aid in material searches for emulation of other systems such as quantum spin glasses or strongly correlated quantum systems for use in error mitigation.

Although we merely studied this system theoretically, this molecule with a different solvent has been recently synthesized and EPR spectroscopy is currently being performed which will be included in future publications. Owing to the relative simplicity of the system, as compared with for instance solid state proposals, and the power of ab initio calculations, we expect a positive correlation between our predicted EPR and that observed experimentally.  

\textit{Acknowledgements.---} This work was supported as part of the
Center for Molecular Magnetic Quantum Materials, an Energy Frontier
Research Center funded by the U.S. Department of Energy, Office of
Science, Basic Energy Sciences under Award No. DE-SC0019330. Computations
were done using the utilities of National Energy Research Scientific
Computing Center, and University of Florida Research Computing systems.

\bibliographystyle{apsrev4-1}

\newpage

\begin{widetext}

\begin{center}
\large{\bf Supplemental Material for ``Majorana Zero Modes Emulated in a Magnetic Molecule Chain'' \\}
\end{center}
\vspace{-6pt}
\begin{center}
Silas Hoffman$^{1}$, Jie-Xiang Yu$^{1}$, Shuang-Long Liu$^{1}$, ChristiAnna Brantley$^{2}$, Guatam D. Stroscio$^{3}$, Ryan G. Hadt$^{3}$, George Christou$^{2}$, Xiao-Guang Zhang$^{1}$, and Hai-Ping Cheng$^{1}$\\
\vspace{2pt}
\it{$^{1}$ Department of Physics, Center for Molecular Magnetic Quantum Materials
and Quantum Theory Project, University of Florida, Gainesville, Florida
32611, USA}
\\
\it{$^{2}$ Department of Chemistry and Center for Molecular Magnetic Quantum Materials,
University of Florida, Gainesville, Florida 32611, USA}
\\
\it{$^{3}$ Division of Chemistry and Chemical Engineering, Arthur Amos Noyes Laboratory of Chemical Physics, California Institute of Technology, Pasadena, California 91125, USA}
\end{center}

\section{Quantum Chemistry Calculations of Monomers} 

We examined a few monomers that may be candidate systems of longer chains for future synthesis. The results are tabulated in Table \ref{tab:casscf}. 

\def\tall{}
\begin{table}[h!]
\begin{center}
\begin{tabular}{c|c|c|c|c|c|c}
\hline
\thead{Material} & \multicolumn{3}{c|}{$D$ ($\textrm{cm}^{-1}$)} & \multicolumn{3}{c}{$E/D$} $\tall$ \\
\hline
& CASSCF & NEVPT2 & Experiment & CASSCF & NEVPT2 & Experiment $\tall$ \\
\hline
Plass Monomers~\cite{SRN909} &  &  &  &  &  & $\tall$ \\
\hline
2-Co1 & -134.6 & -106.6 &  & 0.08 & 0.08 &  $\tall$ \\
2-Co2 & -133.4 & -109.5 &  & 0.11 & 0.11 &  $\tall$ \\
3-Co1 & -128.2 & -109.0 &  & 0.15 & 0.14 &  $\tall$ \\
3-Co2 & -177.8 & -179.9 &  & 0.14 & 0.12 &  $\tall$ \\
\hline
\ce{Co(II)} Monomers &  &  &  &  &  & $\tall$ \\ 
\hline
\ce{[Co(II)](SPh)4]^{2-}}          & -85.2 &  & -70.0~\cite{SLLRN902} & 0.00 &  & 0.09 $\tall$ \\ 
\ce{Co(II)} Schiff base, cn=5 & -33.4 &  & -38.9~\cite{SRN922} & 0.25 &  &      $\tall$ \\ 
\hline
\ce{Ni(II)} Monomers &  &  &  &  &  & $\tall$ \\ 
\hline
$^*$\ce{[Ni(II)(iz)4(ac)2]^{2+}}  &  46.7 &  & -22.3~\cite{SRN923} & 0.28 &  & $\tall$ \\ 
\hline
\end{tabular} 
\end{center}
\caption{\label{tab:casscf}
$D$ and $E$ are the axial and rhombic zero field splitting (ZfS) parameters respectively. CASSCF/NEVPT2 calculations are performed using minimal active space of 3$d$ orbitals.$^*$ 3$d$ orbitals are included in active space (8e,5o). 
}
\end{table}

Our calculations were performed using ORCA version 4.2.1.\cite{SLLRN904, SLLRN905} N-electron valence state second-order perturbation theory (NEVPT2) corrected complete active space self-consistent field (CASSCF) single point calculations~\cite{SLLRN906,SLLRN907,SLLRN908} were performed on all Plass monomers.~\cite{SRN909} Because of the agreement between CASSCF and NEVPT2 values for D and E/D, only CASSCF was performed for the other complexes. The DKH-def2-TZVP~\cite{SRN910} basis set (relativistic recontraction of def2-TZVP~\cite{SRN911}) was used on all atoms. SCF convergence criteria with an energy tolerance of $10^{-7}$ Hartrees were applied. For cobalt complexes, a minimal active space (7e,5o) consisting of the 3d orbitals was used for these calculations ten quartets and forty doublets were included in the state-averaging to account for all the ligand field transitions. For the nickel complex complexes, a minimal active space (8e,5o) consisting of the 3d orbitals was used for these calculations ten triplets and fifteen doublets were included in the state-averaging to account for all the ligand field transitions. 

For DFT optimization of structures, the BP86~\cite{SRN912,SRN913,SRN914} DFT functional was used for structural optimizations. Def2-SVP~\cite{SRN911} and Split-RI-J~\cite{SRN915,SRN916,SRN917,SRN918,SRN919,SRN920}, the default and recommended version of resolution of identities approximation was used; the finest available grids were used (GRID7 NOFINALGRID). Tight SCF convergence criteria, which has a convergence tolerance of $10^{-8}$ Hartrees, was applied for all DFT calculations. The Plass monomers~\cite{SRN909} were optimized using a Zn containing analog to speed SCF convergence similar to as had been done previously. For \ce{[Co(II)(SPh4)]^{2-}}, a previously published, DFT-optimized structure was used.~\cite{SRN921}  The Co(II) Schiff base~\cite{SRN922} and the \ce{[Ni(II)(iz)4(ac)2]^{2+}} structure~\cite{SRN923} were optimized from their crystal structures; in the latter case only hydrogens were optimized. 

\end{widetext}

\section{DFT calculations of additional molecular chains} 

A few Ising crystal systems are reported for which, truncation of long chains into short chains (say 3--10 metal centers) are desirable. In the following, we present DFT calculation results on three such systems, namely $\ce{CoCl2(thiourea)4}$ (nickname DTC), \ce{CoNa(N3)2($\textbf{L}$)}~\cite{C0DT00566E}, and \ce{Co(NCS)2(thiourea)2}~\cite{PhysRevMaterials.5.034401}.

\subsection{DTC}

\begin{figure}[h]
\includegraphics[width=\columnwidth]{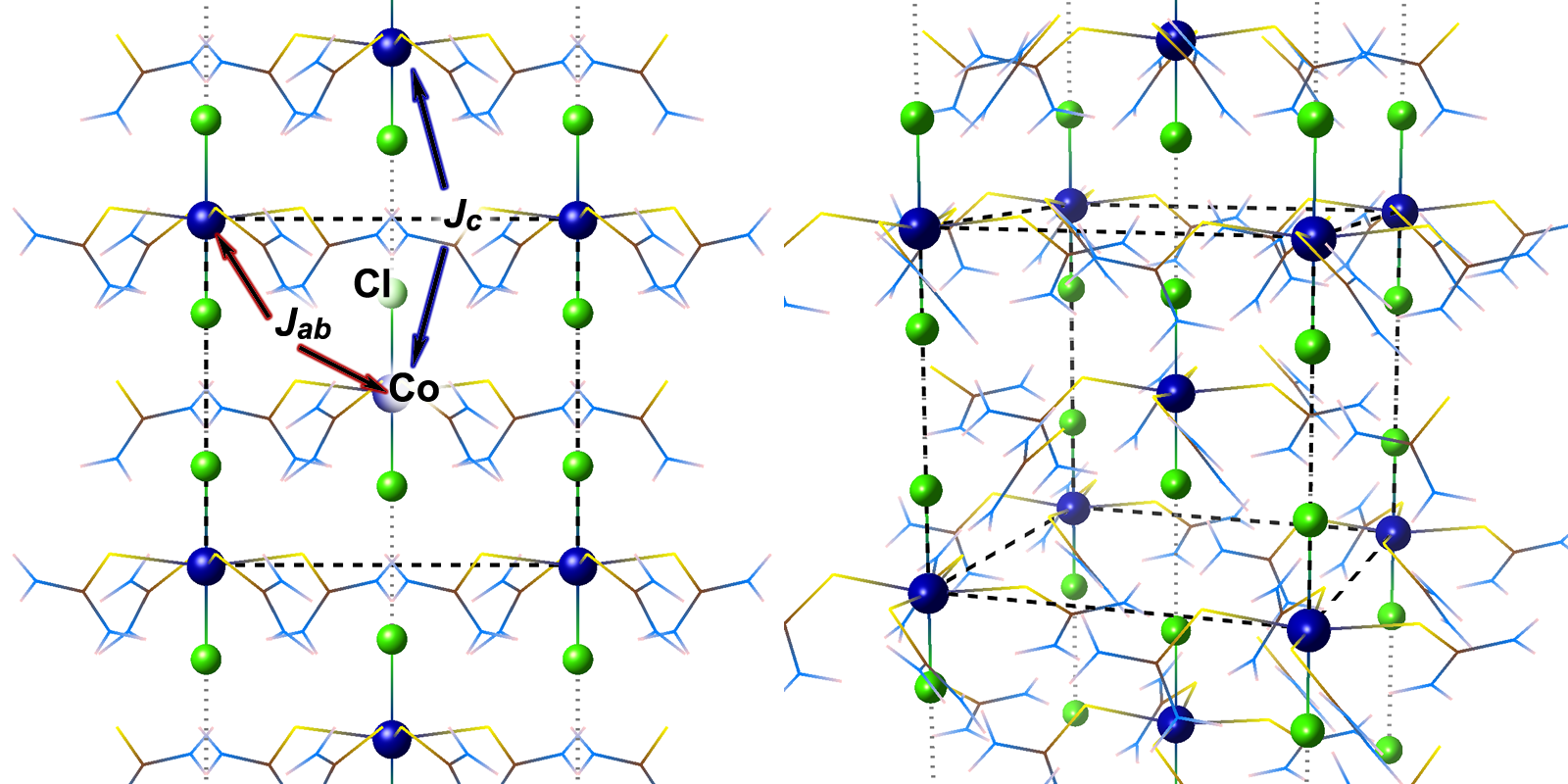} \caption{The structure of molecular crystal DTC in the side view and the perspective
view. Blue and green balls represent Co and Cl atoms, respectively.
The dotted \ce{Cl-Cl} bonds indicate the Van der Waals interactions. The
intra- and inter-chain \ce{Co-Co} interaction $J_{c}$ and $J_{ab}$ are
labeled.}
\label{fig:DTC} 
\end{figure}

Molecular magnets $\mathrm{Co}\mathrm{Cl}_{2}\left(\mathrm{thiourea}\right)_{4}$
(Dichloro-tetrakis(thiourea)-cobalt, DTC, $\mathrm{thiourea}=\mathrm{SC}(\mathrm{NH}_{2})_{2}$)
is the $\mathrm{Co}$ substitution of $\mathrm{Ni}\mathrm{Cl}_{2}\left(\mathrm{thiourea}\right)_{4}$
(Dichloro-tetrakis(thiourea)-nickel, DTN) which is a quasi one-dimensional
quantum molecular magnetic material \cite{JR9630001309,PhysRevB.77.020404,PhysRevB.103.054434}.
DTC is antiferromagnetic materials with Neel temperature 0.92~K and
is suggested to have magnon conductivity below the Neel temperature
\cite{ni_magnon_1971}.

The bulk structure of crystalline DTC has a body-centered tetragonal
lattice where two DTC molecules are in each tetragonal cell. An $1\times1\times2$
supercell with the double size along $c$-axis is used to evaluated
the intra- and inter-chain Co-Co interaction labeled in Fig. \ref{fig:DTC}
based on to the total energy results. By using broken-symmetry method
\cite{Ruiz_1999,Ruiz_2003}, we obtained the intra-chain interaction
$J_{c}=0.18$ meV is ferromagnetic coupling and the inter-chain interaction
$J_{ab}=-0.075$ meV is antiferromagnetic coupling. Similar to DTN,
the strong intra-chain interaction in DTC, is due to the strong Cl-Cl
Van der Waals interactions.

\begin{figure}
\includegraphics[width=\columnwidth]{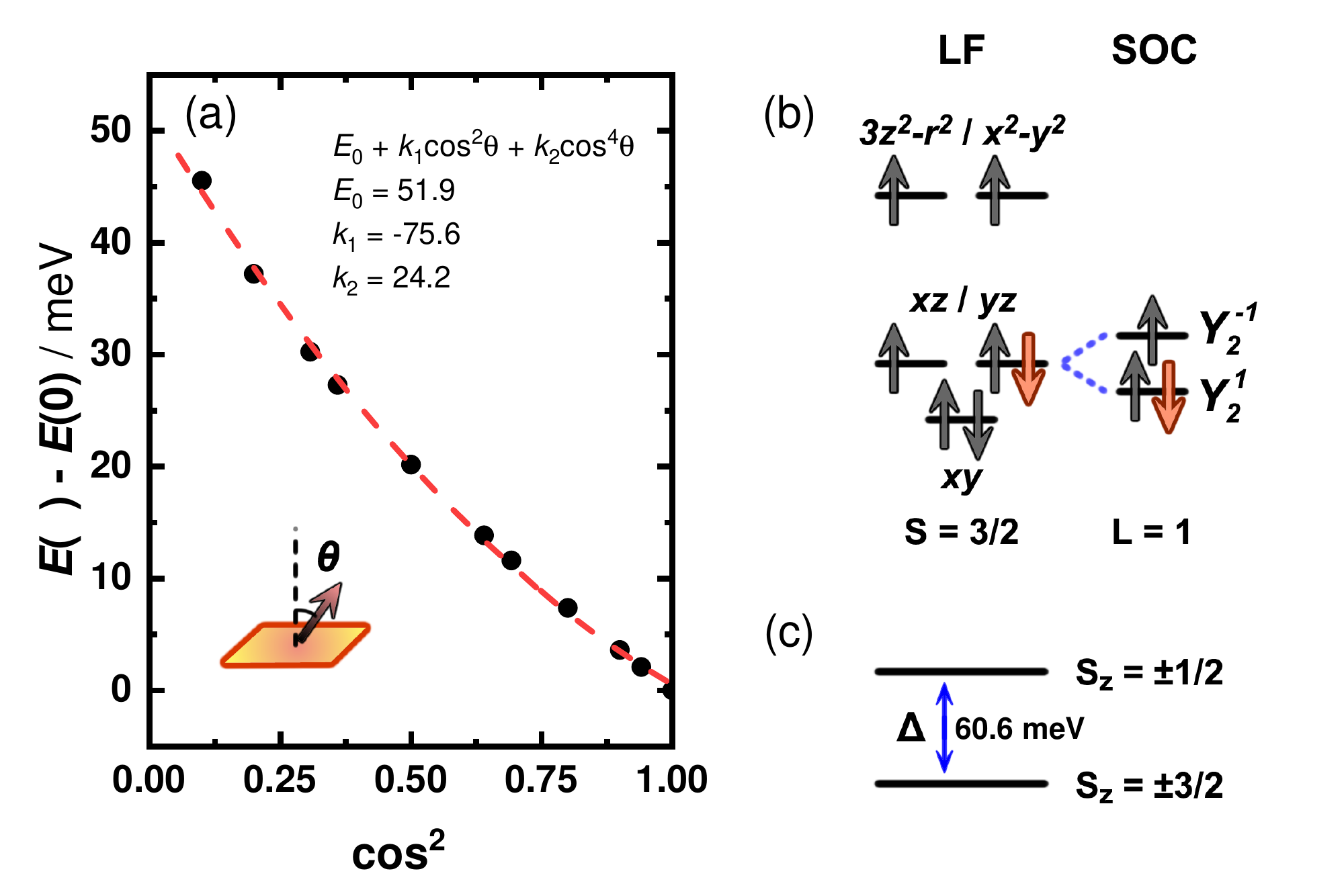} \caption{In DTC, (a) the relative total energy as a function of spin angle
$\cos^{2}\theta$; (b) under tetragonal ligand field (LF) and SOC,
the schematic of electrons in Co($3d$) orbitals which bring about
$S=3/2,L=1$ state; (c) the energy gap $\Delta$ between $S_{z}=\pm3/2$
and $S_{z}=\pm1/2$.}
\label{fig:DTC-zfs} 
\end{figure}

Based on the inter-chain antiferromagnetic state, the lowest energy
state, we included SOC and investigated the magnetic anisotropy by
calculating the total energies with various spin directions. The results
are shown in Fig. \ref{fig:DTC-zfs}(a). The state with out-of-plane
spin directions, corresponding to $\theta=0,\, \cos^{2}\theta=1$ has
the lowest energy, indicating the easy axis is along $c$ axis. The
fitting for $E_{0}+k_{1}\cos^{2}\theta+k_{2}\cos^{4}\theta$ is $k_{1}=-75.6$
meV and $k_{2}=24.2$ meV. The energy difference between $\theta=0$
and $\pi/2$ is about 50 meV, indicating a giant magnetic anisotropy.

The physics is shown in Fig.~\ref{fig:DTC-zfs}(b). The tetragonal
ligand field on Co($3d$) orbitals brings about a partially occupied
two-fold degenerate state dominated by $d_{xz}$ and $d_{yz}$ orbitals.
Due to the SOC, the degenerated $d_{xz}$ and $d_{yz}$ split into
the spherical harmonics $Y_{2}^{1}$ and $Y_{2}^{-1}$. The fully
filled $Y_{2}^{1}$ and the half filled $Y_{2}^{-1}$ provide the
orbital state $L_{z}=1,L=1$. The large orbital momentum along $c$
axis pins the spin to the same direction by $L\cdot S$ and the energy
scale of magnetic anisotropy is $\lambda=21$meV for Co, the coefficient
of SOC, provides the energy scale of magnetic anisotropy. 

\begin{figure}
\includegraphics[width=\columnwidth]{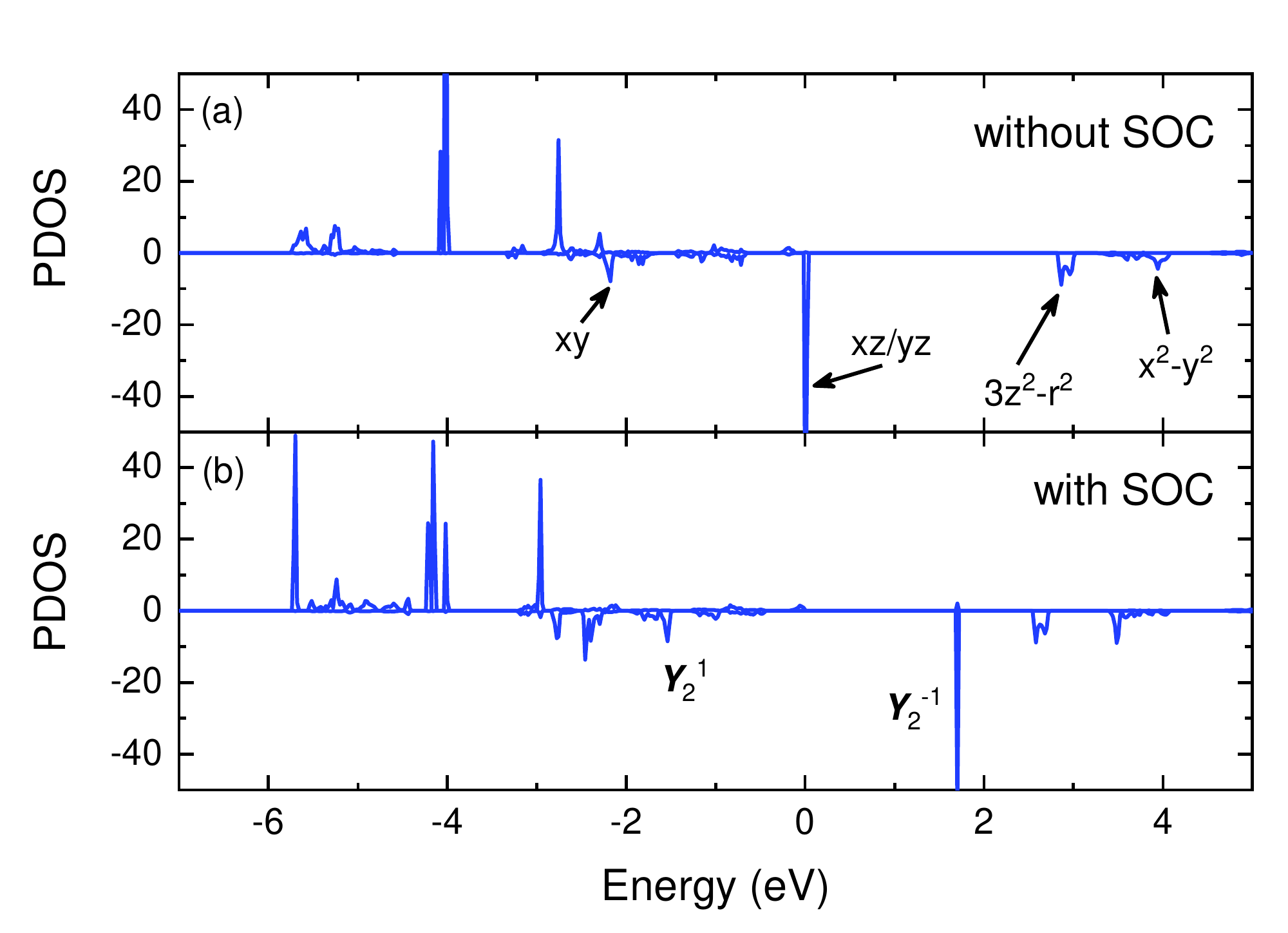} 
\caption{PDOS of Co($3d$) orbitals in DTC (a) without SOC and (b) with SOC.
Positive and nagative values corresponds to the spin-majority and
spin-minority channels, respectively. The Fermi energy is set to zero.} 

\end{figure}

According to projected density-of-state (PDOS) results of Co($3d$)
orbitals, without SOC, the degenerate $d_{xz}$ and $d_{yz}$ orbitals
have a sharp peak at spin-minority channel crossing the Fermi level.
Then SOC lift this degeneracy and open a gap, consisting with the
physics picture. DFT results also show that the orbital moment on
Co is 1.46 $\mu_{B}$ which is even larger than 1.00 $\mu_{B}$ by
$L_{z}=1$. It indicates the hybridization between $d_{xy}$ and $d_{xz}/d_{yz}$.

Note that the coefficient $k_{2}$, the fourth power term is considerable
while for a $S=3/2$ system only the quadratic term $S_{z}^{2}$ is
allowed for the uniaxial magnetic anisotropy. It is because that the
spins with local magnetic moment on Co in DFT calculations is regarded
as a classical vector instead of a quantum spin state. To obtain the
coefficient $\Delta$ of the zero field splitting $-\Delta S_{z}^{2}$,
we converted each classical local magnetic moment state $\alpha$
to the superposition of the quantum spin eigenstates $\left|m\right\rangle $
by $\left|\alpha\right\rangle =\sum_{m=-3/2}^{3/2}c_{m}\left|m\right\rangle $.
Then equation for that that spin configuration is: 
\begin{equation}
E_{\alpha}=E_{0}-\Delta\left\langle \alpha\right|\hat{\mathrm{S}}_{z}^{2}\left|\alpha\right\rangle 
\end{equation}
where $E_{\alpha}$ is the total energy for that spin configuration
and $E_{0}$ is the spin-irrelevant energy. By solving the linear
equations, we evaluated $\Delta=60.6$ meV. $\Delta$ is also the energy
gap $\Delta$ between two quantum spin states $S_{z}=\pm1/2$ and
$S_{z}=\pm3/2$ (Fig. \ref{fig:DTC-zfs}(c)). Since $\Delta$ is much
large than $J_{c}$ as well as the energy scale of thermal fluctuation,
the spin states on DTC's Co always lay in $S_{z}=\pm3/2$.

The shortcoming of DTC is that the system is not a covalently bonded spin chain although the anisortropy is huge. However, it can be used as a guidance for future synthesis. 

\subsection{DFT results on \texorpdfstring{$\mathrm{Co}\mathrm{Na}\left(\mathrm{N}_{3}\right)_{2}\left(\mathbf{L}\right)$}{Lg}}

\begin{figure}[ht]
\includegraphics[width=\columnwidth]{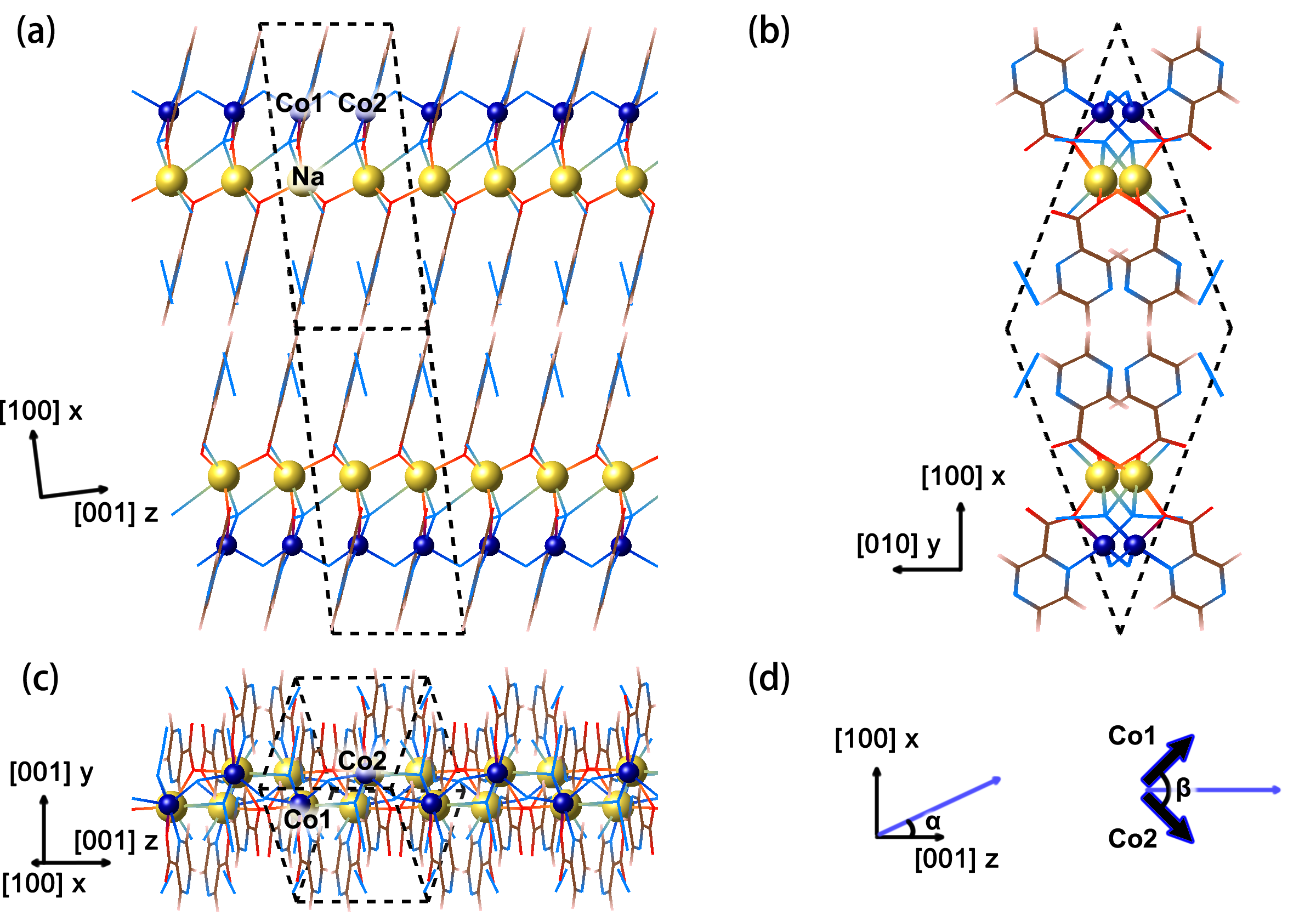}
\caption{(a)-(c) three views of the structure of crystalline $\mathrm{Co}\mathrm{Na}\left(\mathrm{N}_{3}\right)_{2}\left(\mathbf{L}\right)$. Two Co atoms, Co1 and Co2, are labeled in the unit cell. (d) $\alpha$ is the angle between the total spin direction and $\left[001\right]$ direction and $\beta$ is the angle between $\mathbf{S}_{1}$ and $\mathbf{S}_{2}$, the spins on Co1 and Co2 respectively, representing the non-collinear spins. }
\label{fig:CoNaN3L}
\end{figure}

$\mathrm{Co}$-azido compounds $\mathrm{Co}\mathrm{Na}\left(\mathrm{N}_{3}\right)_{2}\left(\mathbf{L}\right)$, (\textbf{L} = Pyrazine-2-carboxylato) is also a single-chain magnet.~\cite{C0DT00566E} 
Its crystalline phase has a based-centered monoclinic lattice with space group $C2/c$ (No. 15), shown in Fig. \ref{fig:CoNaN3L}. Each chain has two $\mathrm{Co}$ atoms, labeled Co1 and Co2, in one unit cell.

We perform DFT calculations with and without spin-orbit coupling included to investigate the magnetic properties of $\mathrm{Co}\mathrm{Na}\left(\mathrm{N}_{3}\right)_{2}\left(\mathbf{L}\right)$.
In the calculations, the $ab$-plane in Fig.~\ref{fig:CoNaN3L} is set to the $xy$-plane so that $\left[001\right]$ or $z$ direction has an angle $6.94^{\circ}$ with the $c$-axis. 

We first perform total energy calculations for collinear spin configurations on $\mathrm{Co}$. 
The total energy results are shown in Table. \ref{tab:CoNaN3L-energy}.
Without spin-orbit coupling (SOC), 
the energy difference between ferromagnetic (FM) and antiferromagnetic (AF) spin configurations per Co-Co bonds $\Delta E_{AF-FM}$ is 4.64 meV, indicating the FM coupling. 
When SOC is included, $\Delta E_{AF-FM}$ is 4.88 meV, almost invariant with spin directions, indicating that the anisotropic exchange interaction is ignorable. 
Therefore, the magnitude of exchange parameter $J$ is $\Delta E_{AF-FM}/2 S_{1} S_{2}= 1.08$~meV (using broken-symmetry method: $J=\Delta E_{AF-FM}/\left(2 S_{1} S_{2}+S_{1\left(2\right)}\right)= 0.81$~meV), where $S_{1}=S_{2}=3/2$ is the spin of Co1 and Co2 respectively.

\begin{table}
\caption{In $\mathrm{Co}\mathrm{Na}\left(\mathrm{N}_{3}\right)_{2}\left(\mathbf{L}\right)$, the total energy results for all collinear spin-configurations, including $\Delta E_{AF-FM}$, the energy difference between ferromagnetic (FM) and antiferromagnetic (AF) spin configurations per Co-Co bonds, and the total energies relative to $\left[102\right]$ (The direction with the lowest total energy) per Co. non-SOC refers to the results without SOC included. }
\begin{tabular}{crr}
\hline
 & $\Delta E_{\mathrm{AF-FM}}$ (meV) & $E -  E_{\left[102\right]}$ (meV) \\ 
\hline
        without SOC & 4.640  & -- \\
$\left[001\right]$  &  4.879 &  0.287 \\
$\left[010\right]$  &  4.873 &  0.118  \\
$\left[100\right]$  &  4.900 &  0.635 \\
$\left[011\right]$  &  4.879 &  0.197 \\
$\left[101\right]$  &  4.897 &  0.017 \\
$\left[110\right]$  &  4.869 &  0.408 \\
$\left[0\bar{1}1\right]$  &  4.880 & 0.196 \\
$\left[10\bar{1}\right]$  &  4.884 &  0.922 \\
$\left[1\bar{1}0\right]$  &  4.884 &  0.393 \\
$\left[111\right]$  & -- & 0.059 \\
$\left[\bar{1}11\right]$  & -- &  0.645 \\ 
$\left[1\bar{1}1\right]$  & -- &  0.058 \\
$\left[11\bar{1}\right]$  & -- &  0.647 \\
$\left[102\right]$  & -- &  0.000 \\
$\left[10\bar{2}\right]$  & -- &  0.724 \\
$\left[201\right]$  & -- &  0.200 \\
$\left[20\bar{1}\right]$  & -- &  0.937 \\
\hline
\end{tabular}
    \label{tab:CoNaN3L-energy}
\end{table}

\begin{figure}[ht]
\includegraphics[width=\columnwidth]{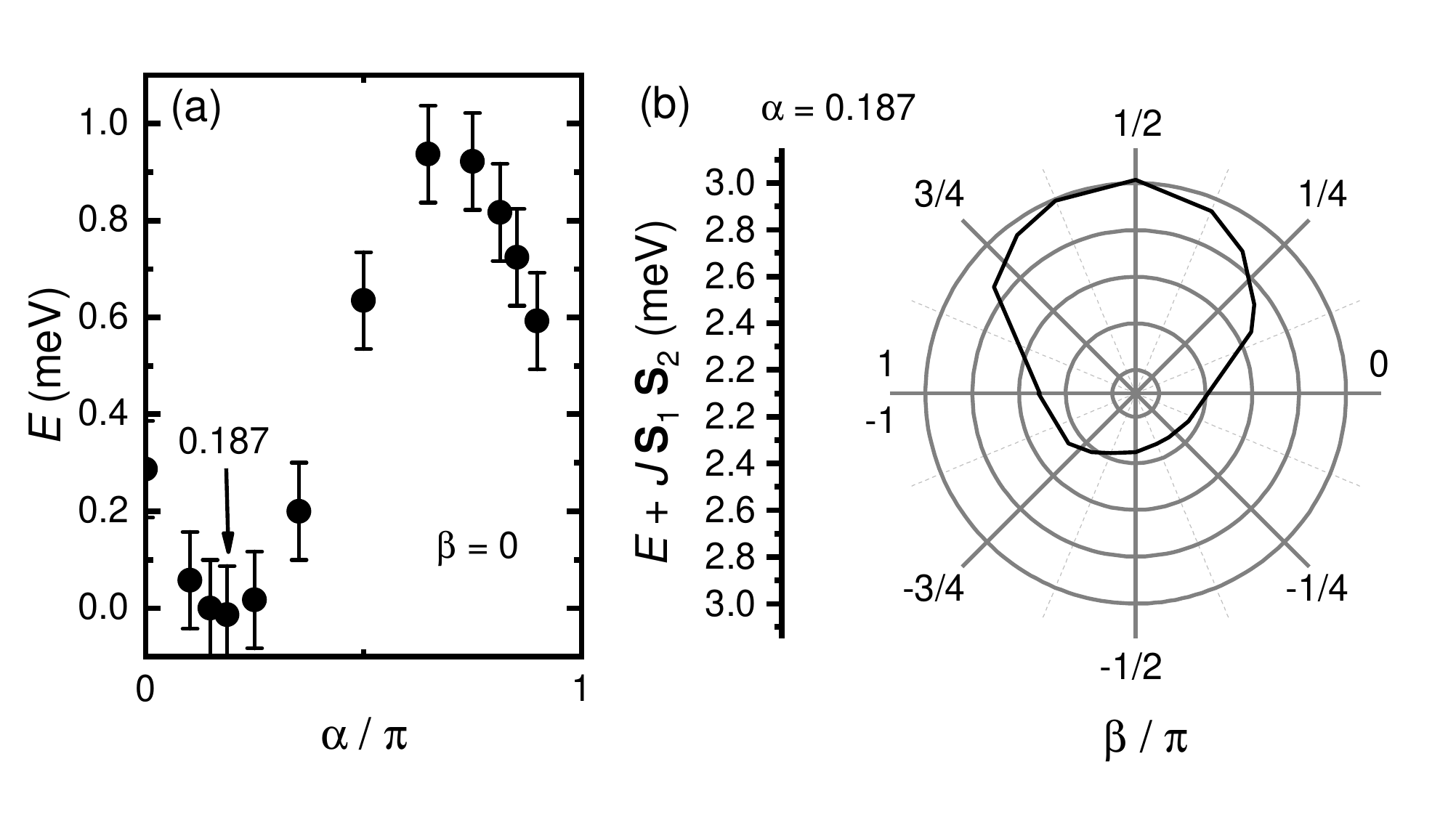}
\caption{(a) The relative total energy $E$ for collinear spin configurations ($\beta=0$) as a function of total magnetization angle $\alpha$. The error bar as $\pm0.1 $meV refers to the precision of DFT. (b) The energy $E+J\mathbf{S}_{1}\cdot\mathbf{S}_{2}$ ($J=1.08$~meV) for non-collinear spin alignments as a function of non-collinear angle $\beta$. The magnetization angle $\alpha$ is fixed at $0.187\pi$.}
\label{fig:CoNaN3L-MAE}
\end{figure}

The two neighboring $\mathrm{Co}$ atoms in one chain have the same magnitude and direction of magnetic anisotropy along the $x$ and $z$ directions, but they have opposite sign in magnetic anisotropy along the $y$ direction. To determine the magnitude and direction of magnetic anisotropy, we selected the total energy results for collinear FM spin configurations in $xz$-plane, as a function of $\alpha$, the angle between total magnetization and $\left[001\right]$ direction, shown in Fig.~\ref{fig:CoNaN3L-MAE}(a).
The energy minimum is at $\alpha=1.87\pi$ and the maximum energy difference is $0.92$ meV.

Then we set non-collinear spin configurations where the total spins $\mathbf{S}_{1}+\mathbf{S}_{2}$ is along $\alpha=1.48\pi$ and the angle between $\mathbf{S}_{1}$ and $\mathbf{S}_{2}$ is $\beta$.
The contribution of exchange interaction $-J \mathbf{S_{1}}\cdot\mathbf{S}_{2}$ is subtracted from the total energies and the results are shown in Fig.~\ref{fig:CoNaN3L-MAE}(b). 
The energy minima are at $\beta= -0.30\pi$ and $1.70\pi$ and the maximum energy difference is $0.68$ meV.
Considering the on-site anisotropy Hamiltonian $\mathcal{H}_{aniso}=-D \tilde{S}_{z}^2+E\left(\tilde{S}_{x}^2-\tilde{S}_{y}^2\right)$ where the direction of $\tilde{S}_{z}$ is the easy axis, we can obtain the magnitude of $D$ as $0.36$~meV (quamtum: $0.54$~meV) and $E$ as $0.05$~meV (quamtum: $0.07$~meV).

\subsection{\texorpdfstring{\ce{Co(NCS)2(thiourea)2}}{Lg}}

\begin{figure}[ht]
\includegraphics[width=\columnwidth]{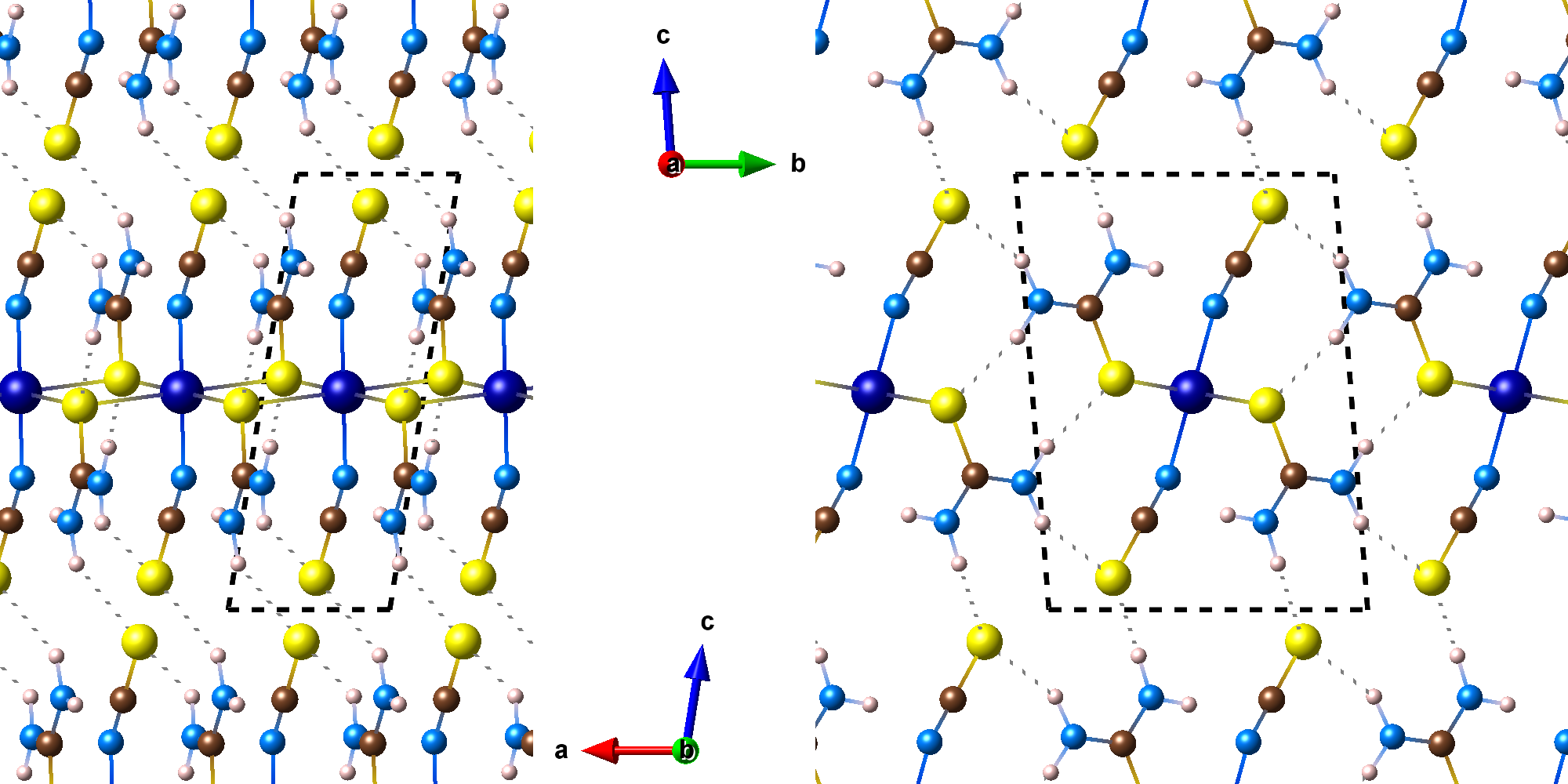}
\caption{The structure of crystalline $\mathrm{Co}\left(\mathrm{NCS}\right)_{2}\left(\mathrm{thiourea}\right)_{2}$. Dotted \ce{S-H} bonds are H-bonds.} 
\label{fig:CoNCST}
\end{figure}

$\mathrm{Co}\left(\mathrm{NCS}\right)_{2}\left(\mathrm{thiourea}\right)_{2}$ ($\mathrm{thiourea} = \mathrm{SC}(\mathrm{NH}_{2})_{2}$) is a new synthesised single-chain magnet (SCM) \cite{PhysRevMaterials.5.034401}. 
It is a $\mathrm{Co}$-centered SCM where each $\mathrm{Co}$ atom with $S=3/2$ high-spin state is surrounded by four in-plane $\mathrm{S}$ atoms of ligands $\mathrm{thiourea}$ and two out-of-plane $\mathrm{N}$ atoms of ligands $\mathrm{NCS}$ 
so that each $\mathrm{Co}$ is in a octahedral ligand-field, shown in Fig.\ref{fig:CoNCST}. 
Its crystalline phase has a triclinic lattice with space group $\mathrm{P}\bar{1}$ so that it only has a inversion center on the centered $\mathrm{Co}$ atom.

According to the experiment, the intra-chain $\mathrm{Co}$-$\mathrm{Co}$ spin exchange interaction is ferromagnetic coupling with $J\sim0.31$ meV  ($3.63$ K) and the inter-chain exchange interaction is very weak antiferromagnetic coupling with  $J'\sim-0.027$ meV ($-0.31$ K). Each $\mathrm{Co}$ has a very strong uniaxial magnetic anisotropy with $D\sim8.6$ meV ($100$ K) along \ce{N-Co-N} direction. Our preliminary DFT results do not match with the experimental findings. Further DFT calculations are on-going. 

\section{Ising model}

The conventional starting point is the $N$ site spin chain with Ising axis
along the $x$ direction in a transverse magnetic field, 
\begin{equation}
H=-J\sum_{i=1}^{N-1}\sigma_{i}^{x}\sigma_{i+1}^{x}-h\sum_{i=1}^{N}\sigma_{i}^{z}\,,
\label{ising}
\end{equation}
where $\sigma_{i}^{x}$ ($\sigma_{i}^{z}$) rotates spin $i$ about
the $x$ ($y$) axis. $J$ is the exchange strength and $h$ is the
magnitude of the magnetic field. When $h>J$ the system is paramagnetic
with all spins pointing along the direction of the magnetic field.
When $h<J$, the system is a degenerate ferromagnet. In particular,
when the magnetic field is zero, the degenerate ground states are
the configurations of the spins uniformly pointing parallel or uniformly
pointing antiparallel to the Ising axis.

Using these spin operators one can construct a Jordan-Wigner (spinless)
fermion at each site according to $c_{j}=(\sigma_{j}^{x}-i\sigma_{j}^{y})[\prod_{l<j}(-\sigma_{l}^{z})]/2$.
Inserting this definition into Eq.~(\ref{ising}), the fermionic
system is described by 
\begin{equation}
H=-J\sum_{i=1}^{N-1}\left(c_{i}^{\dagger}c_{i+1}+c_{i}^{\dagger}c_{i+1}^{\dagger}+\textrm{H.c}\right)-2h\sum_{i=1}^{N}(c_{i}^{\dagger}c_{i}-1/2)\,.\label{kit}
\end{equation}
which is simply the Kitaev model with nearest neighbor hopping and
superconducting pairing $J$, chemical potential $2h$. In fermionic
description, the system is topological trivial when $h>J$ and topologically
nontrivially when $h<J$. According to the bulk-edge correspondence,
this implies that in the latter case, there exists localized MBSs
at the ends. In the Ising case, these take particularly simple forms
in terms of the spin operators, $\gamma'=\sigma_{1}^{x}$ and $\gamma=-i\sigma_{N}^{x}\prod_{N}(-\sigma_{l}^{z})$.
These can be formed into a fermionic operator, $f=\sigma_{1}^{x}+\sigma_{N}^{x}\prod_{N}(-\sigma_{l}^{z})/\sqrt{2}$,
which also commuates with the Hamiltonian. Thus, the ground states
can be characterized by the occupancy of this fermion, $\left|1\right\rangle=(\left|\Uparrow\right\rangle+\left|\Downarrow\right\rangle)/\sqrt{2}$
and $\left|0\right\rangle=(\left|\Uparrow\right\rangle-\left|\Downarrow\right\rangle)/\sqrt{2}$.

Since the EPR will ultimately describe the transitions between energy
levels, let us take a closer look at spectra in both the magnetic
and fermionic pictures. Finding the eigenvalues of Eq.~(\ref{ising})
gives the energies of all the states in the system, $E_{n}$. Because
Eq.~(\ref{kit}) is quadratic in the fermionic creation and annihilation
operators, we can write it as a tight-binding model. Thus, the spectrum
in the fermionic picture describes the set of levels, $\mathcal{E}_{n}$
that can be occupied or unoccupied and the sum of the occupied levels
describes the total energy of the system. Excited states are created
by moving an occupied state to an unoccupied state; this is stipulated
by the fact that states with energy below the chemical potential (hole
states) must be moved to their particle conjugate. 

We wish to subject our spin chain to a transverse oscillating magnetic field
described by, 
\begin{equation}
H_{\textrm{EPR}}(t)=\cos(\Omega t)\sum_{i=1}^{N}\sigma_{i}^{z}\sim\cos(\Omega t)\sum_{i=1}^{N}c_{i}^{\dagger}c_{i}\,.
\end{equation}
The linear response of the system in the spin language is \cite{berimJETP79,kuboJoPSoJ54}
\begin{equation}
G(\Omega)\sim\int dt\cos(\Omega t)\left\langle \left\{ \sum_{i=1}^{N}\sigma_{i}^{z},\sum_{i=1}^{N}\sigma_{i}^{z}(t)\right\} \right\rangle \,,
\label{respA}
\end{equation}
where the expectation is with respect to one of the ground states
and $\sigma_{z}(t)$ evolves according to Eq~(\ref{ising}). For
the spin system, this is straightforward to calculate numerically
as we can directly evaluate the integrand for a sufficiently small
system and, upon integration, we pick out the states with phase $\Omega t$.

In the fermionic language we require a bit more
massaging before we can numericaly compute the EPR response. First,
note that $\sigma_{i}^{z}=2c_{i}^{\dagger}c_{i}-1=c_{i}^{\dagger}c_{i}-c_{i}c_{i}^{\dagger}$.
For superconducting Hamiltonians, it is typical to use the Nambu doubling
procedure in which we define a vector $\vec{c}_{i}=(c_{i},c_{i}^{\dagger})$.
Notice that this makes Eq.~(\ref{kit}) easy to handle in a single
particle formalism. Also, $\sigma_{i}^{z}=\vec{c}_{i}^{\dagger}\tau^{z}\vec{c}_{i}$
with $\tau^{z}$ the Pauli $z$ matrix acting in Nambu space. Let
me now write yet another vector $\vec{C}=(c_{1},c_{1}^{\dagger},\ldots,c_{N},c_{N}^{\dagger})$
which is $2N$-dimensional. The total spin along the $z$ axis can
then be written as $\vec{C}^{\dagger}(\mathbb{I}_{N\times N}\otimes\tau^{z})\vec{C}=\sum_{\alpha,\beta=1}^{2N}C_{\alpha}^{\dagger}(\tau^{z})_{\alpha\beta}C_{\beta}$
where we have dropped the identity matrix. Let $\Lambda_{\nu\beta}$
be a matrix of the eigenvectors of Eq.~(\ref{kit}) in rows so that
the eigenoperators are $\vec{D}=\hat{\Lambda}\vec{C}$ or $D_{\nu}=\sum_{\beta}\Lambda_{\nu\beta}C_{\beta})$.
Here, $[H,D_{\nu}]=\omega_{\nu}D_{\nu}$ and $D_{\nu}|\textrm{gnd}\rangle=0$
if $E_{\nu}>0$. Thus, the expression for the response in the fermionic
language is 
\begin{widetext} 
\begin{equation}
G(\Omega)\sim\int dt\cos(\Omega t)\sum\left[\left\langle D_{\mu'}^{\dagger}\Lambda_{\mu'\alpha'}\tau_{\alpha'\beta'}^{z}(\Lambda^{\dagger})_{\beta'\nu'}D_{\nu}'e^{-i\omega_{\mu}t}D_{\mu}^{\dagger}\Lambda_{\mu\alpha}\tau_{\alpha\beta}^{z}(\Lambda^{\dagger})_{\beta\nu}D_{\nu}e^{-i\omega_{\nu}t}\right\rangle +\textrm{H.c.}\right]\,,
\end{equation}
where the summation is over repeated indices. Let us focus on the
expectation value. 
\begin{align*}
 & \sum\left\langle D_{\mu'}^{\dagger}\Lambda_{\mu'\alpha'}\tau_{\alpha'\beta'}^{z}(\Lambda^{\dagger})_{\beta'\nu'}D_{\nu}'e^{-i\omega_{\mu}t}D_{\mu}^{\dagger}\Lambda_{\mu\alpha}\tau_{\alpha\beta}^{z}(\Lambda^{\dagger})_{\beta\nu}D_{\nu}e^{-i\omega_{\nu}t}\right\rangle \\
 =& \sum \left\langle \Lambda_{\mu'\alpha'}\tau_{\alpha'\beta'}^{z}(\Lambda^{\dagger})_{\beta'\nu'}\Lambda_{\mu\alpha}\tau_{\alpha\beta}^{z}(\Lambda^{\dagger})_{\beta\nu}\right\rangle e^{-i(\omega_{\nu}-\omega_{\mu})t}\left\langle D_{\mu'}^{\dagger}D_{\nu'}D_{\mu}^{\dagger}D_{\nu}\right\rangle \\
 =& \sum \left\langle \Lambda_{\mu'\alpha'}\tau_{\alpha'\beta'}^{z}(\Lambda^{\dagger})_{\beta'\nu'}\Lambda_{\mu\alpha}\tau_{\alpha\beta}^{z}(\Lambda^{\dagger})_{\beta\nu}\right\rangle e^{-i(\omega_{\nu}-\omega_{\mu})t}(\langle D_{\mu'}^{\dagger}D_{\nu'}\rangle\langle D_{\mu}^{\dagger}D_{\nu}\rangle+\langle D_{\mu'}^{\dagger}D_{\nu}\rangle\langle D_{\nu'}D_{\mu}^{\dagger}\rangle)\\
 =& \sum \left\langle \Lambda_{\mu'\alpha'}\tau_{\alpha'\beta'}^{z}(\Lambda^{\dagger})_{\beta'\nu'}\Lambda_{\mu\alpha}\tau_{\alpha\beta}^{z}(\Lambda^{\dagger})_{\beta\nu}\right\rangle e^{-i(\omega_{\nu}-\omega_{\mu})t} \left(\delta_{\mu'\nu'}\delta_{\mu\nu}\Theta(-E_{\mu'})\Theta(-E_{\nu}) +  \delta_{\mu'\nu}\delta_{\nu'\mu}\Theta(E_{\mu})\Theta(-E_{\nu}) \right)\\
 =& \sum\Theta(-E_{\mu'})\Theta(-E_{\mu})\Lambda_{\mu'\alpha'}\tau_{\alpha'\beta'}^{z}(\Lambda^{\dagger})_{\beta'\mu'}\Lambda_{\mu\alpha}\tau_{\alpha\beta}^{z}(\Lambda^{\dagger})_{\beta\mu}+ e^{-i(\omega_{\mu}'-\omega_{\mu})t}\Theta(E_{\mu})\Theta(-E_{\mu'})\Lambda_{\mu'\alpha'}\tau_{\alpha'\beta'}^{z}(\Lambda^{\dagger})_{\beta'\mu}\Lambda_{\mu\alpha}\tau_{\alpha\beta}^{z}(\Lambda^{\dagger})_{\beta\mu'}\,,
\end{align*}
\end{widetext}
where the Heaviside theta function is due to the occupancy (vacancy)
of the states below (above) the Fermi energy. This quantity can be
readily evaluated thus the response calculated. From the form of these
expressions, the second term excites a filled state to a higher energy
state then returns the system to the ground state. The first term
is just the square of expectation value of the magnetization. 

Although the magnetic case allows us to treat a more general system,
i.e. when the effective fermion parity is not conserved, the fermionic
language allows us to treat a considerably larger chain as the dimension
of the Hamiltonian in the magnetic and fermionic pictures is $2^{N}$
and $2N$, respectively. As we have checked to ensure both pictures
result in identical EPR response, we present the results using the
fermionic picture unless otherwise specified.

At this point we remark that the spin system is inversion symmetric,
i.e. $i\rightarrow N-i+1$. That is, in the fermion language, the
inversion operator is $\hat{\mathcal{I}}=\sum_{j=1}^{N}c_{j}^{\dagger}c_{N-j-1}$
and $[\hat{\mathcal{I}},H]=0$. Moreover, because $\hat{\mathcal{I}}^{2}=\mathbb{I}$,
the eigenvectors of the Hamiltonian are also eigenvectors of $\hat{\mathcal{I}}$
with eigenvalues of $\pm1$. Lastly we note that $\hat{\mathcal{O}}=\sum_{j=1}^{N}c_{j}^{\dagger}c_{j}$
commutes with $\hat{\mathcal{I}}$. In first quantized language the
latter is simply $\tau_{\alpha\beta}^{z}$ and $\Lambda_{\mu\alpha}\tau_{\alpha\beta}^{z}(\Lambda^{\dagger})_{\beta\mu'}$
is matrix element of $\hat{\mathcal{O}}$ with respect to the eigenvectors
$\mu$ and $\mu'$. However, because $\hat{\mathcal{O}}$ cannot mix
eigenvectors with different eigenvalues of $\hat{\mathcal{I}}$, EPR
can any excite any filled state to half (rounded down) of the excited
particle states. This will be important one analyzing the EPR spectra.

\begin{figure}[ht]
\includegraphics[width=\columnwidth]{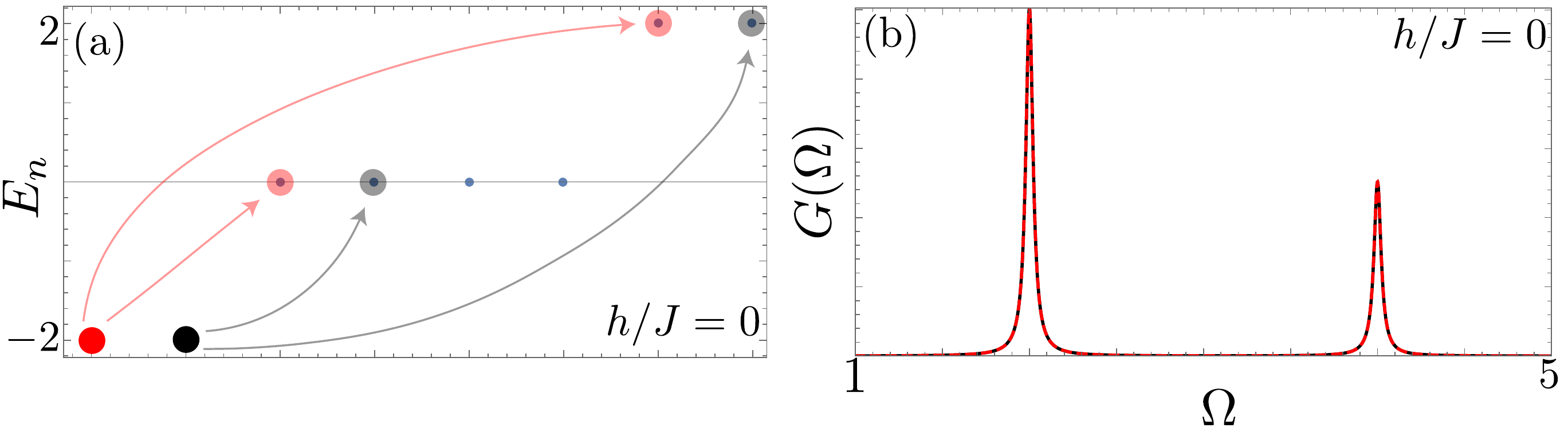} \caption{Spectrum indicating the transitions (a) that result in the EPR signal as a function of ac-magnetic field frequency, $\Omega$ (b), with $h=0$.}
\label{EPR1} 
\end{figure}

\begin{figure}[ht]
\includegraphics[width=\columnwidth]{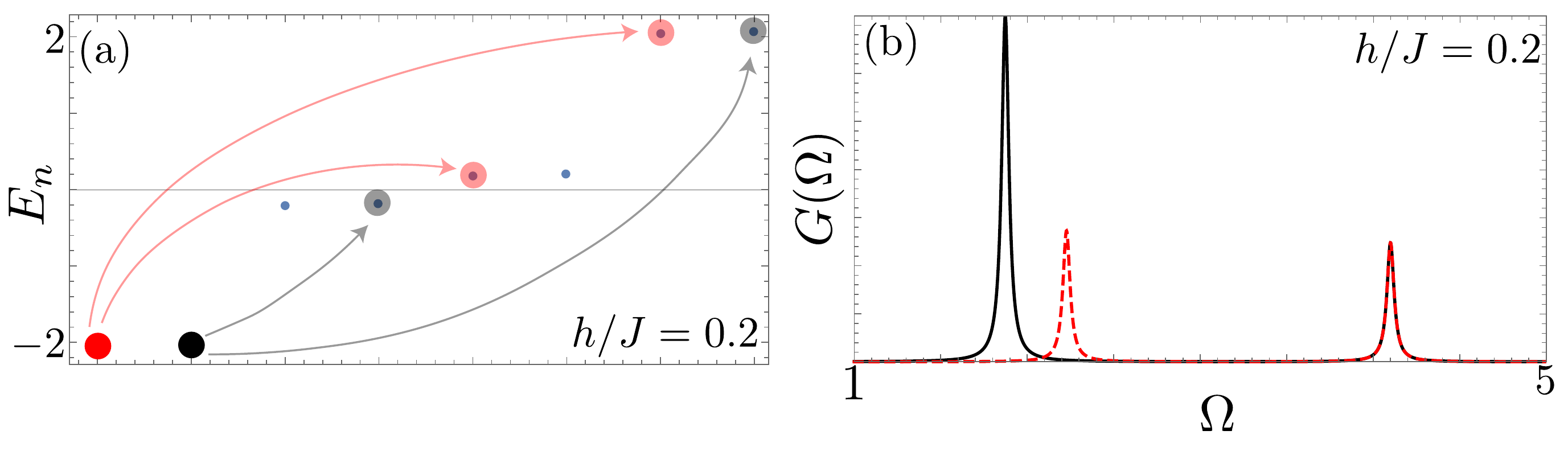} \caption{Spectrum indicating the transitions (a) that result in the EPR signal as a function of ac-magnetic field frequency, $\Omega$ (b), with $h/J=0.2$.}
\label{EPR2} 
\end{figure}

We calculate the EPR spectrum of a three-site chain in zero magnetic
field with respect to both ground states (Fig.~\ref{EPR1}). The
low energy peak corresponds to exciting the zero energy state to an
excited state while the high energy peak corresponds to exciting quasiparticle
hole. Because the quasiparticle energies are degenerate, there is
no difference in the spectra between the ground states. In finite
magnetic field, the band is no longer flat. Moreover, because the
zero energy states have opposite eigenvalues of the operator $\hat{\mathcal{I}}$,
the EPR spectrum of the ground states is different (Fig~\ref{EPR2}).

\bibliographystyle{apsrev4-1}

\end{document}